\documentclass[aps,prb,twocolumn,floatfix,showpacs]{revtex4-2}

\usepackage{graphicx}
\usepackage{amsmath,amssymb}
\usepackage{dcolumn} 
\usepackage{bm} 
\usepackage{xcolor}
\usepackage[utf8]{inputenc}

\usepackage{hyperref}
\hypersetup{%
  colorlinks=true,
  citecolor=blue,
  urlcolor=blue,
  linkcolor=blue,
  }

\begin{document}


\title{Laser-induced ultrafast insulator--metal transition in  $\text{BaBiO}_{3}$ }

\author{Alexander E. Lukyanov$^{1,2}$}
\author{Vyacheslav D. Neverov$^{1,2}$}
\author{Yaroslav V. Zhumagulov$^{3,2,1}$}
\author{Alexey P. Menushenkov$^{1}$}
\author{Andrey V. Krasavin$^{1}$}
\author{Alexei Vagov$^{2}$}

\affiliation{$^{1}$National Research Nuclear University MEPhI, Kashirskoye shosse 31, Moscow, 115409, Russian Federation}
\affiliation{$^{2}$ITMO University, St. Petersburg 197101, Russia}
\affiliation{$^{3}$University of Regensburg, Regensburg, 93040, Germany}

\date{\today}
\begin{abstract}
We investigate ultra-fast coherent quantum dynamics of undoped  $\text{BaBiO}_{3}$  driven by a strong laser pulse. Our calculations demonstrate that in a wide range of radiation frequencies and intensities the system undergoes a transient change from the insulating to the metallic state, where the charge density wave and the corresponding energy spectrum gap vanish. The transition takes place on the ultra-fast time scale of tens femtoseconds, comparable to the period of the corresponding lattice vibrations. The dynamics are determined by a complex interplay of the particle-hole excitation over the gap and of the tunnelling through it, giving rise to the highly non-trivial time evolution which comprises high harmonics and reveals periodic reappearance of the gap. The time evolution is obtained by solving the dynamical mean-field theory equations with the realistic parameters for the system and radiation. Results are summarized in the phase diagram, helpful for a possible experimental setup to achieve a dynamical control over the conduction state of this and other materials with the similarly strong electron-phonon interaction.
\end{abstract}

\maketitle

\section{Introduction}

Perovskite compounds based on $\text{BaBiO}_{\text{3}}$ (BBO) are a special class of the high-temperature superconductors (HTSC) that do not include transition metal ions. As a rule, such structures have a complex phase diagram with many structural phases that depend on the doping level as well as on the temperature. These include dielectric, metallic and superconductive phases, with the critical temperature of the superconductive transition varying between $T_c = 13 K$ for the led doped materials \cite{Sleight1975} and $T_c = 34 K$ for the potassium doping \cite{PhysRevB.37.3745}. 

Unique properties of those materials are related to their peculiar crystal structure and its dynamical distortions (phonons), and to the Fermi surface configurations.  For example, when doped with potassium with the concentration $x \geq 0.37$, the local electronic structure is characterised by a spatially separated mixture of free electrons and localized electron pairs \cite{Menushenkov2001, Menushenkov2002, Menushenkov2016}. The electron pairs determine the charge transfer and formation of the superconductive state, while free electron states play a role in the insulator-metal phase transition and the appearance of the Fermi-liquid state above the percolation threshold \cite{Menushenkov2000}.  

\begin{figure}
\begin{center}
    \includegraphics[height=3.1cm]{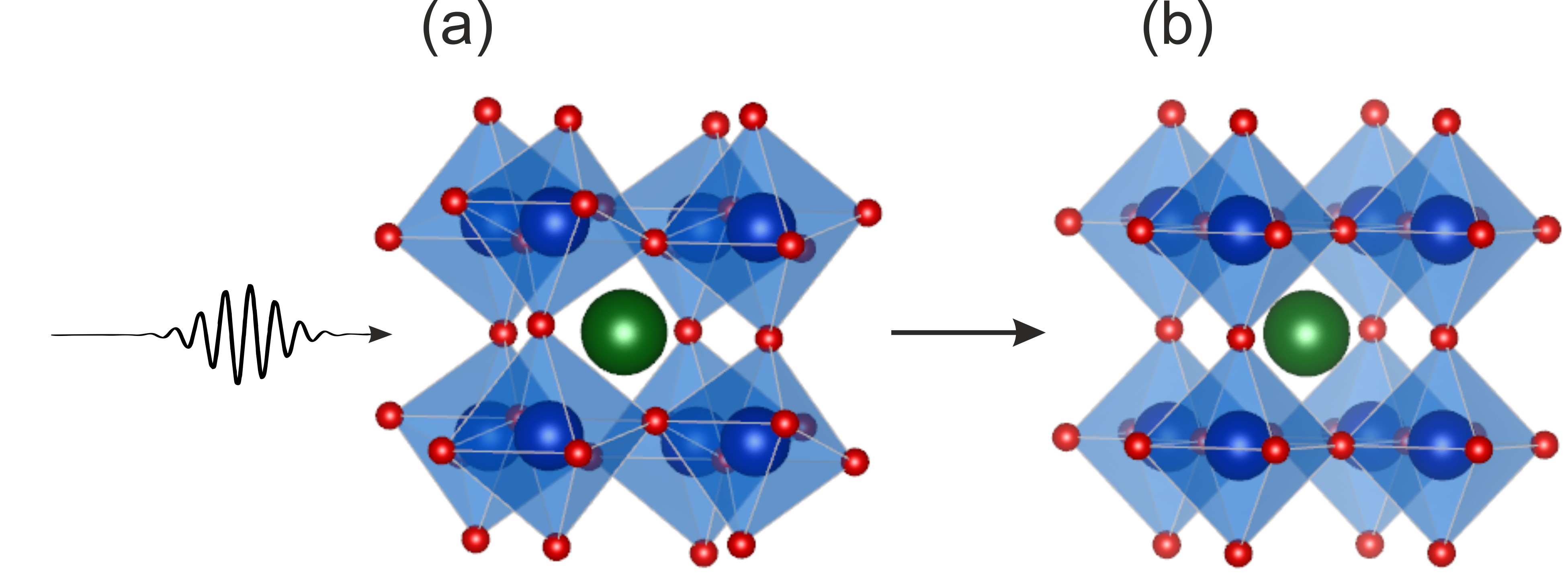}
\end{center}
\caption{A schematic crystal structure of $\text{BaBiO}_{\text{3}}$, with ions $\text{Bi}$ (blue), $\text{O}$ (red)  and $\text{Ba}$ (green). (a) The structure before the pulse, distorted by the CDW, has larger and smaller $\text{BiO}_{\text{6}}$ octahedra, extra electron/hole pairs sit on the upper antibonding $\text{Bi6s-O2p}_{\sigma ^{\ast}}$ orbitals. (b) The laser excitation breaks the pairs on the antibonding orbitals and removes the lattice distortion.}
 \label{fig:PumpExp}
\end{figure}

Bismuthates are characterized by strong deformations of the crystal lattice associated with the anomalously high amplitude of their phonon oscillations \cite{Uchida1987}.  In this context the most interesting is the ''breathing'' phonon mode with alternating sizes of the $\text{BiO}_{\text{6}}$ octahedra and $\text{Bi--O}$ bonding along three crystallographic directions \cite{SLEIGHT2015152, PhysRevLett.117.037002} (see Fig. \ref{fig:PumpExp}). The size alternation involves the charge transfer by electron pairs tunnelling between the two neighbouring octahedra. The octahedron with the extra electron pair is larger, whereas the one without it is smaller. This migration of electron pairs can be interpreted as the ''valence disproportionation'' where $\text{Bi}^{\text{3+}}$ and $\text{Bi}^{\text{5+}}$ ions are formed instead of two neighbouring  $\text{Bi}^{\text{4+}}$ ions \cite{PhysRevLett.61.2713,PhysRevB.76.174103}. Different electronic filling of the $\text{Bi6s-O2p}_{\sigma ^{\ast}}$ orbitals of the neighboring $\text{BiO}_{\text{6}}$ octahedra creates a dynamical double-well potential for the vibration of oxygen ions, experimentally observed in EXAFS spectra \cite{Menushenkov1998, Menushenkov2000}.
 
The inherently strong coupling between electrons in the conduction band and lattice distortions profoundly affects the carriers state in those materials. Specifically, in the parent compound, $\text{BaBiO}_{\text{3}}$ with the Fermi level in the middle of the conduction band this coupling gives rise to the charge density wave (CDW) accompanied by the optical gap in the single-particle electronic spectra of the system, estimated in various studies as being $\simeq 2$eV  \cite{PhysRevB.42.923, Tajima1985, Uchida1987, Sato1989, Tang2007}. 

Details of the electronic spectra as well as of the carrier-phonon coupling in those materials can be probed by studying their properties when the system is externally driven from its equilibrium, creating the system dynamical response.  For example, the gap structure of the spectrum is investigated by the pump-and-probe measurements of the optical conductivity (see \cite{Nicoletti2017,Buzzi2018}), where the charge carriers are excited above the gap leading to a transient metallic state,  gradually relaxing towards the original insulating state by the electron-hole recombination. The system can be driven so strongly that its original phase is transiently changed.  

An adequate interpretation of such experiments requires modelling the dynamics far from the equilibrium. This has to be done beyond the single-particle picture since the system state is altered considerably. In addition, many relevant non-equilibrium phenomena take place on the ultra-fast time scale, where quantum coherences play a large role and cannot be neglected.  

Sufficiently accurate and consistent analysis of the ultra-fast dynamic in strongly correlated systems, performed on equal footing with the analysis of their equilibrium state, became possible relatively lately, owing to rapid advances in the computing power. The dynamics of strongly-correlated systems far from equilibrium have been addressed in a number of recent works, that studied, in particular, electron-mediated relaxation after ultra-fast pumping \cite{Moritz2013}, dynamical redistribution of spectra \cite{Kemper2014}, the role of the phonon-carrier coupling in the time-resolved optical spectra \cite{Sentef2013},  dynamics of the photodoped charge transfer insulators \cite{Gole2019}, the gap dynamics in excitonic insulators \cite{Mor2017} and the dynamics of a periodically driven strongly-correlated Fermi-Hubbard model \cite{Sandholzer2019}. 

The BBO is a representative example of the systems where the insulating CDW state appears as the result of the strong phonon-carrier coupling. Whether the CDW in this material can be externally manipulated and the metallic state achieved transiently by applying a strong driving is a very interesting problem, which until now has been investigated neither experimentally nor theoretically. Generally, little is known about the dynamics of this structure when it is driven far from equilibrium.

In this work, we investigate the ultrafast dynamics of the $\text{BaBiO}_{\text{3}}$ excited by a strong continuous laser pulse. In our analysis we consider a time evolution of the system retaining all pertinent quantum coherences within the dynamical mean-field approach adapted to study strongly correlated systems \cite{Georges1996,Kotliar2006}. The chosen level of the theory is sufficient to describe both the metal-insulator transition of the system and its ultra-fast time evolution on the equal footing,  ensuring consistency of the results.

Our analysis demonstrates that a sufficiently strong excitation can break the insulating state and the system evolves into the metallic phase on the time scale of tens femtoseconds. It is shown that the dynamics of this regime is highly non-linear, comprises many high harmonics and even develops the continuous high-frequency spectra. Although our study focuses on  $\text{BaBiO}_{\text{3}}$, the results are general and can be used as a guidance for experimental setups to control the state of the system with the strong electron-phonon interaction.

The paper is organized as follows. In Section \ref{Model} we introduce the model, describe the used assumptions, present necessary details of the formalism and discuss the choice of the system parameters. Results of the calculations are found in Section \ref{Results} in which we give examples of the various types of the dynamical behaviour of the system as the reaction to the applied excitation field, and give a summary of the obtained results in the form of the phase diagram. Finally, Section \ref{Conclusions} briefly summarizes the results of the work and outlines future perspectives.

\begin{figure}
    \centering
    \includegraphics[width=0.25\textwidth]{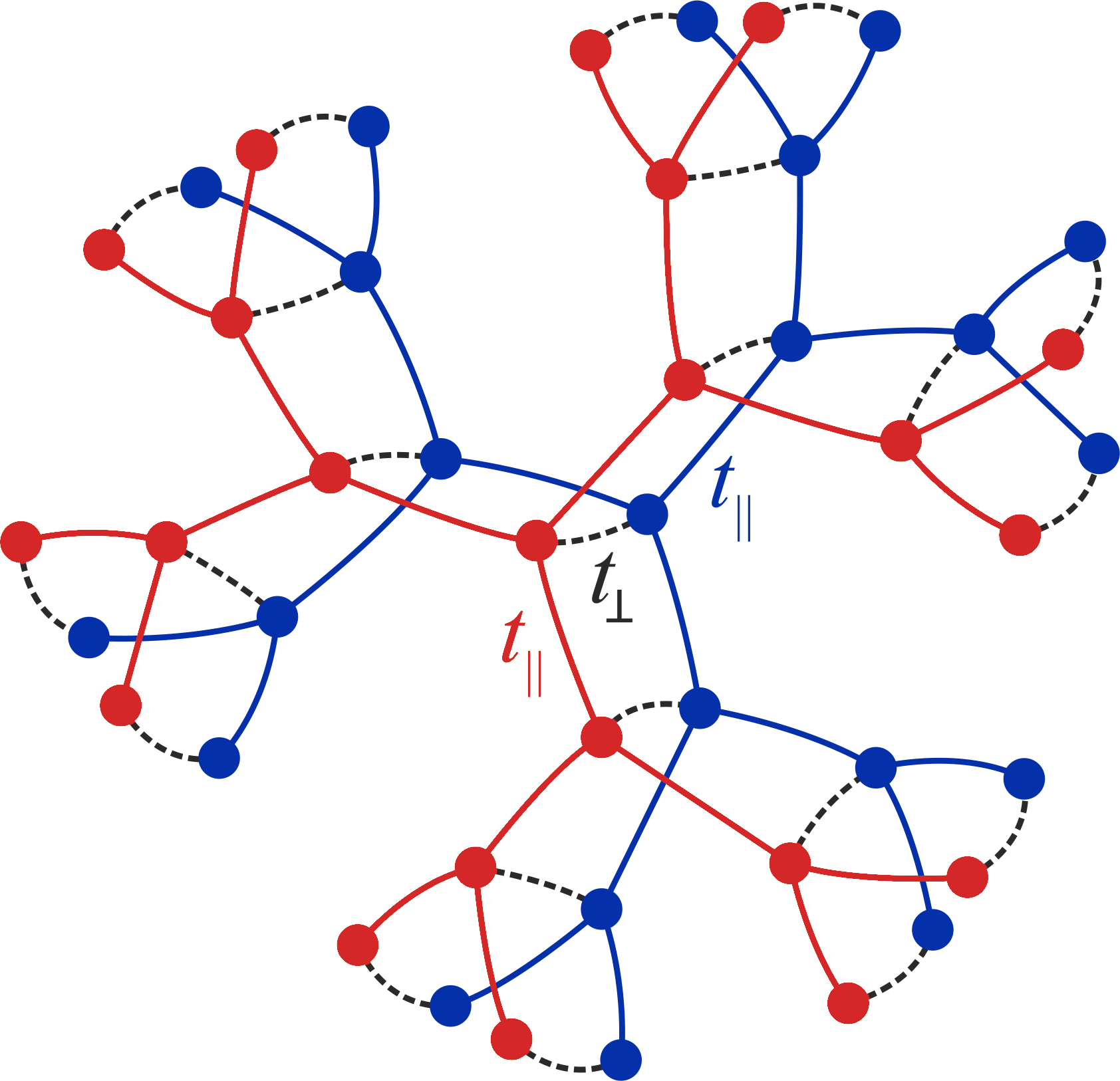}
    \caption{Double Bethe lattice. Solid and dashed lines correspond to hoppings $t_{\|}$ within and $t_{\perp}$ between the lattices, respectively. Each lattice is shown with coordination number 3 for clarity.}
    \label{fig:Lattice}
\end{figure}

\section{Model and Methods} \label{Model}

\subsection{Holstein model for $\text{BaBiO}_{\text{3}}$}

To describe the metal-insulator  transition in $\text{BaBiO}_{\text{3}}$ we adopt the tight-biding (TB) model for electrons that occupy  $\text{BiO}_{\text{6}}$ octahedra and can tunnel between them. The octahedra are regarded as "sites" and the hopping elements describe the tunnelling. The coupling to the lattice deformation modes is taken into account within the Holstein model \cite{Khazraie2018} where an electron of an octahedron is coupled to the distortion oscillation (phonon) mode of this octahedron.  This model neglects the interaction between phonon modes of different octahedra. The Hamiltonian of the Holstein model is  
\begin{align} 
\label{H_TB}
\begin{split}
H =  -\sum_{ij \sigma} \left( t_{ij} c^\dagger_{j \sigma}c_{i \sigma} +\text{h.c.} \right)  -\mu \sum_i (n_{i\uparrow} + n_{i\downarrow})  \\ +g\sum_i \left(b^\dagger_i+b_i \right) \left(n_{i\uparrow} + n_{i\downarrow} - 1 \right)
 +\omega_{\textrm{ph}}\sum_i b^\dagger_i b_i,
\end{split}
\end{align}
where $c_{i\sigma}$ denotes the electron operators with spin $\sigma$ located at site $i$,  $n_{i\sigma} = c_{i\sigma}^\dagger c_{i\sigma}$ is the electron number operator, $\mu$ is the chemical potential, $t_{ij}$ are the hopping integrals,  $b_i$ is a phonon operator corresponding to the deformation mode on site $i$,  $\omega_{\textrm{ph}}$ is the mode frequency and $g$ is the phonon-electron coupling constant. Values of $t_{ij}$ are extracted from the {\it ab initio} calculations by comparing obtained single-particle density of states (DOS) with the results for the tight-biding model (\ref{H_TB}). The comparison yields that the tight-binding model needs to account for the nearest as well as the second- and the fourth-nearest neighbours. The corresponding hopping matrix elements are extracted as $t_{1}=-0.45$eV, $t_{2}=-0.09$eV, $t_{4}=0.10$eV, respectively, while the chemical potential is $\mu=-0.13$eV.

\subsection{Bethe lattice model}

The Holstein model (\ref{H_TB}) appears simple, however, is still quite difficult to solve, even within the dynamical mean-field theory (DMFT). However, one can simplify the calculations considerably by taking into account the details of the metal-insulator transition. The transition is associated with the CDW state where the size of the neighbouring octahedra $\text{BiO}_{\text{6}}$ alternate. The CDW state of interest is associated with the doubling of the unit cell of the simple cubic lattice of BaBiO$_3$ in all three directions so that the corresponding CDW vector is $(\pi, \pi, \pi)$. This choice extracted from  experimental observations for BaBiO$_3$ \cite{Tajima1987,Menushenkov2000,Harris2020}. 

\begin{figure}[t!]
\begin{center}
    \includegraphics[width=0.5\textwidth]{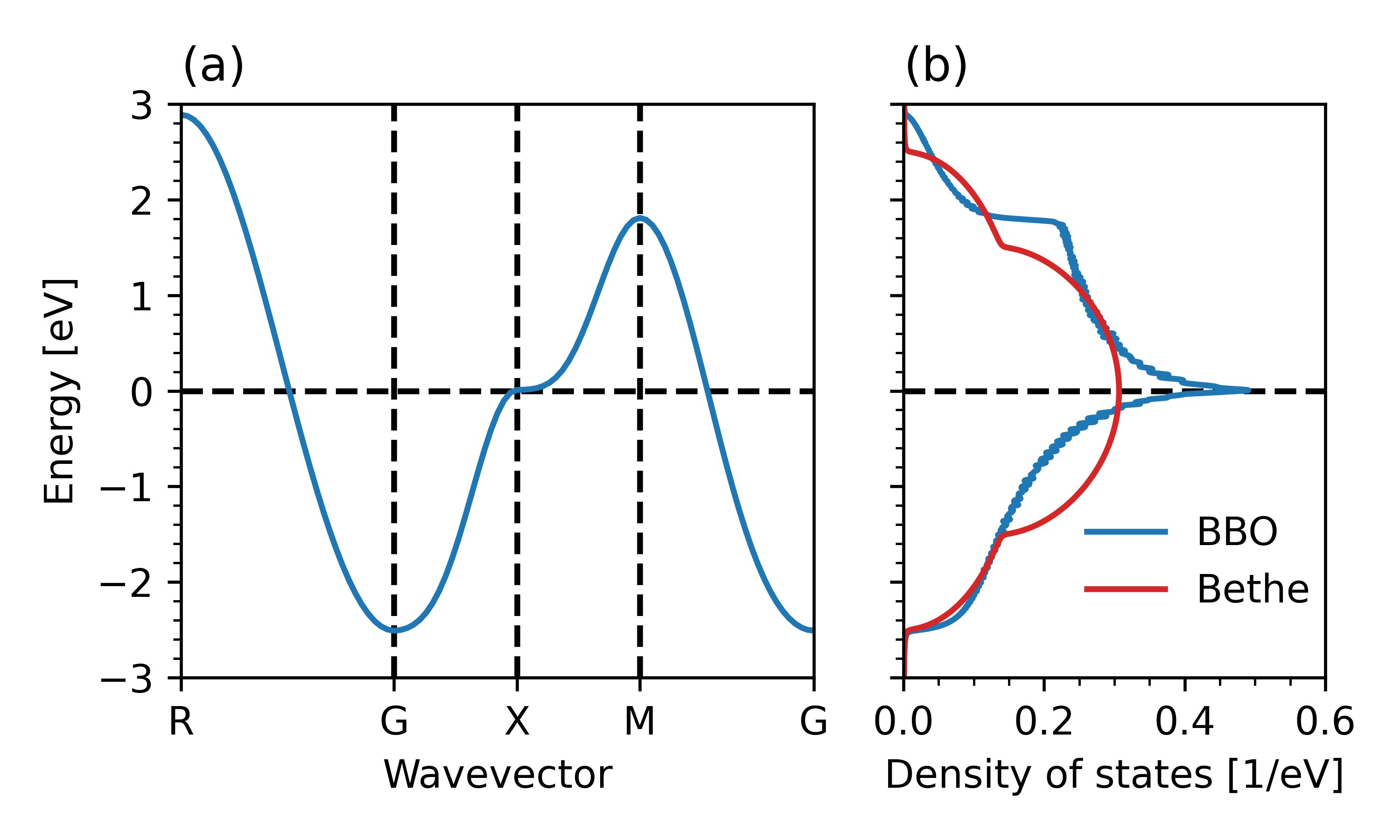}
\end{center}
\caption{(a) Band structure of the tight-binding model (\ref{H_TB}) for the cubic BBO structure \cite{Khazraie2018}.   (b) Single particle DOS calculated for the original TB model with the cubic BBO structure  (blue) and for the double Bethe lattice model (red).}
\label{fig:dos}
\end{figure}

The alternation of the neighbouring sites and the appearance of the CDW state can be conveniently described by separating the original system into two sublattices of the double-size elementary cell, each containing octahedra of the same state.  Within this model, the metal-insulator transition takes place when electronic states of the sublattices become different. The dynamics within sublattices is described using the Bethe lattice model, schematically illustrated in Fig. \ref{fig:Lattice} for the coordinate number 3.  It is effectively infinite-dimensional model with a continuum of electronic states which gives rise to the quantum relaxation kinetics. The main advantage of the Bethe model is that its DMFT description yields exact results. It is commonly assumed that the Bethe lattice model offers a good approximation for 3D systems \cite{Georges1996} and can be used as a trusted approximation to investigate dynamical properties of many-body systems (see, e.g., \cite{Kotliar2006, Ciuchi1999, Moeller1999, Hafermann2009, Harland2020}, and also \cite{Harland2016} and references therein). Here it is adopted to describe the dynamics of each sublattice. At the same time the model with two Bethe sublattices is adequate to describe a CDW state with the broken symmetry between sublattices.

For the $\text{BaBiO}_{\text{3}}$ structure of interest our final model consists of two equivalent Bethe lattices (see Fig. \ref{fig:Lattice}) where the electron can hope between the nearest neighbours of each sublattice (intra-lattice hopping) as well as between the equivalent sites of different sublattices (inter-lattice hopping). In the following, we refer to this model as the double Bethe lattice model (DBLM).
The Hamiltonian of the model reads as 
\begin{align} \label{H}
H = & -t_{\|}\sum_{\langle i,j \rangle \alpha \sigma} c^\dagger_{j\alpha \sigma}c_{i\alpha \sigma} - t_{\perp} \sum_{i \sigma}(c^\dagger_{i 0 \sigma}c_{i 1 \sigma}+\text{h.c.} )  \notag \\ & -\mu\sum_{i \alpha \sigma} n_{i\alpha \sigma}  
+g\sum_{i\alpha} \left(b^\dagger_{i\alpha} + b_{i\alpha} \right) \left(n_{i\alpha \uparrow} + n_{i\alpha \downarrow} - 1 \right) \notag \\
&+\omega_{\textrm{ph}}\sum_{i\alpha} b^\dagger_{i\alpha} b_{i\alpha},
\end{align}
where $\alpha=0,1$ is the sublattice index, the sum with $\langle i,j \rangle$ goes over the nearest neighbour sites,  $t_{\|}$ and $t_{\perp}$  denote the intra- and inter-lattice hopping matrix elements, respectively. Their values are chosen such that the model yields the best possible approximation for the DOS. Figure \ref{fig:dos} shows the band structure (Panel a) and the DOS (Panel b) of the TB model as well as the DOS of the DBLM. The best fit of the DOS is achieved for $t_{\|}=1$eV and $t_{\perp}=0.5$eV. The Bethe lattice gives a qualitatively similar DOS but it does not capture the van Hove singularity in the middle of the band.  The van Hove singularity facilitates the instability towards the formation of the CDW state when the Fermi level is close to it, $\mu \simeq 0$.  However, the DBLM demonstrates this instability without the singularity in the DOS. 

Notice that the dynamics of electrons with different spins as determined by the Hamiltonian (\ref{H}) is independent and can be obtained by considering the spinless version of the model (\ref{H}), where the the carrier density $n_{i \alpha}$ enters the phonon coupling as $n_{i \alpha \uparrow} + n_{i \alpha \downarrow} = 2n_{i \alpha}$.

Finally, the coupling between the laser field and the system carriers is taken into account via the Peierls substitution, where the hopping integrals are assigned with the phase as $t_{ij} \to t_{ij} \exp( i \varphi_{ij})$, calculated as the integral of the vector potential $\varphi_{ij} = (e/c \hbar ) \int_{i}^j ({\bf A} \cdot d {\bf r})$ between the sites $i$ and $j$. In the calculations we assume that the external driving affects only the coupling between the sublattices with the corresponding hopping matrix elements
\begin{align} 
\label{Peierls}
t_{\perp}(t) = t_{\perp} e^{i \varphi (t)}, \quad \varphi (t)=f(t) \sin (\omega_0 t),
\end{align}
where $\omega_0$ is the frequency of the driving oscillations and $f(t)$ is the envelope function of the pulse. The approximation where the  driving affects only the coupling between sublattices follows from the goal of our study that is to describe the non-linear dynamics and the relaxation of the CDW state when the system is subject to laser pulses of high intensity. In the chosen model the laser excites dynamical degrees of freedom  associated with the CDW. We note, however, that in the regime of strongly non-linear dynamics of interest in this work all dynamical degrees of freedom (excitation modes), including those describing the dynamics within sublattices, are intermixed. Therefore, all such modes will be eventually excited regardless of the driving type. Finally, in this work we investigate the case of the constant driving so that the pulse arrives at time $t=0$ and then has a constant amplitude,  $f(t) = \Theta (t) A_0$, where $A_0$ is the dimensionless driving strength and $\Theta(t) = 1$ at $t>0$ and zero otherwise.

\begin{figure}
    \centering
    \includegraphics[width=0.4\textwidth]{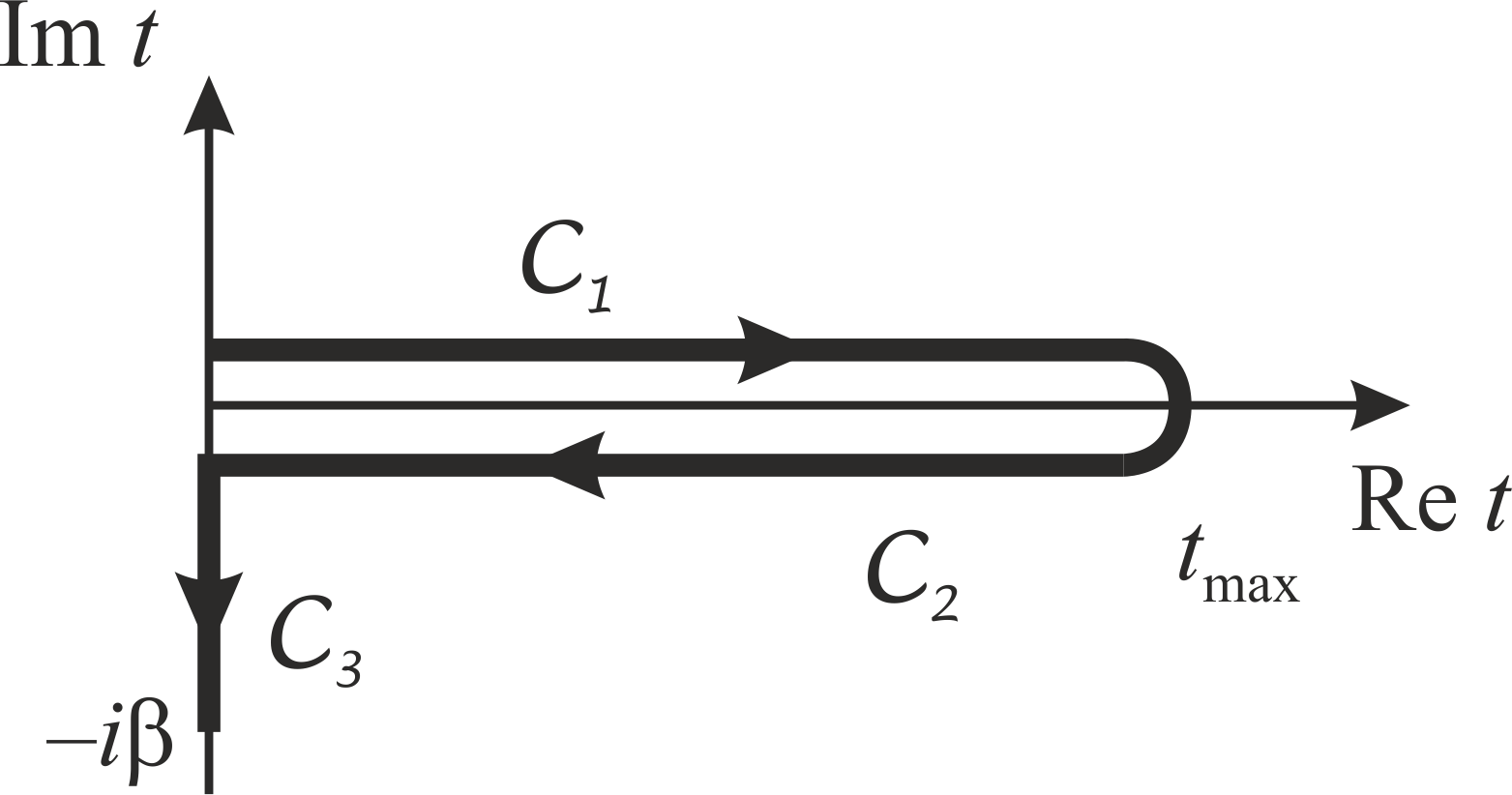}
    \caption{Kadanoff-Baym contour with three time domains: real-time forward $\mathcal{C}_1$ and backward $\mathcal{C}_2$, and imaginary-time $\mathcal{C}_3$. The arrows set the contour time ordering; $t_\textrm{max}$ corresponds to the maximum real-time propagation of the solution. }
    \label{fig:Contour}
\end{figure}

\subsection{DMFT for the Bethe lattice model}

To calculate the system dynamics we employ the DMFT approach \cite{Georges1996}, which maps the original many-body problem to a model with an effective impurity. Its application to the model with a single Bethe lattice has a number of crucial simplifications with respect to the original TB model \cite{Moeller1999}. Here we extend this approach to the DBLM in Eq. (\ref{H}), where the impurity becomes a dimer, defined on both sublattices. The main DMFT equations in this case retain their general appearance, however, acquire a form of the 2$\times$2 matrix equations for the corresponding matrix Green functions. For the convenience of the reader we briefly outline the main equations for the dynamics together with the used approximations.   

All pertinent physical quantities of the system are found from the two-time electron $G(t,t')$ and phonon $D(t,t')$  matrix Green functions for the dimer impurity, which are obtained by solving the Dyson equations
\begin{align}
&D(t,t')=D_0(t,t')+\int_{\mathcal{C}}d{ \tau }d \tau^\prime \: D_0(t, \tau ) \Pi ( \tau , \tau^\prime ) D(\tau^\prime ,t'), \notag \\
&\left[ i \partial_t - \epsilon(t) \right]G(t,t')-\int_{\mathcal{C}}d{\tau} \: {\Sigma}({\tau},t)G(t,{\tau})=\delta_{\mathcal{C}}(t,t'), 
\label{Dyson}
\end{align}
where the integration is performed along the Kadanoff--Baym (KB) contour $\mathcal{C}$ in Fig. \ref{fig:Contour} and  $\delta_{\mathcal{C}}(t,t')$ is the $2\times 2$ matrix delta function defined on this time contour.  The phonon bare Green function is defined by
\begin{align} \label{D0}
D^{-1}_0(t,t')=-\frac{1}{2\omega_{\textrm{ph}}} \left( \partial^2_t + \omega^2_{\textrm{ph}} \right)  \delta_\mathcal{C}(t,t'),
\end{align}
and $\epsilon(t)$ is the one-particle Hamiltonian,
\begin{align} \label{epsilon}
 \epsilon(t)=\begin{pmatrix} \mu+\Sigma^{MF}_0(t) & t_{\perp}(t) \\ t_{\perp}^\ast(t) & \mu+\Sigma^{MF}_1(t) \end{pmatrix}
\end{align}
with the mean-field contribution to the self-energy 
\begin{align} \label{MF}
\Sigma^{\textrm{MF}}_\alpha(t)=g \bar{X}_\alpha(t), 
\end{align}
where the dimensionless quantity $\bar{X}_\alpha(t) =  X_\alpha(t) \sqrt{2M \omega_{\textrm{ph}}/\hbar}$ ($M$ is the oxygen mass) related to the sublattice displacement $X_\alpha(t)$, which measures the structural distortion of the octahedra and is calculated by averaging the phonon operators. The impurity self-energy for electrons in Eq. (\ref{Dyson}) comprises two components,
\begin{align} \label{Sigma_imp}
{\Sigma} (t,t') =\Sigma_{\textrm{ME}} (t,t') + \Delta_G (t,t'),
\end{align}
where the hybridization function $\Delta_G(t,t')$ in the case of the Bethe lattice has the form \cite{Georges1996}
\begin{align} \label{Delta}
\Delta_G(t,t')=t^2_{\|}G(t,t'),
\end{align}
whereas for the contribution $\Sigma_{\textrm{ME}}$ to the self-energy beyond the mean-field we use the self-consistent Migdal approximation \cite{Murakami2015, Murakami2016}
\begin{align} 
\label{Sigma_sMig}
\Sigma_{\textrm{ME}}(t,t')=ig^2 D(t,t')G(t,t').
\end{align}
For the the phonon impurity self-energy this approximation yields 
\begin{align} \label{Pi_sMig}
\Pi(t,t') \rightarrow \Pi_{\textrm{ME}}(t,t')=-2ig^2 G(t,t')G(t',t).
\end{align}

Solutions to the Dyson equations are used to obtain the average lattice displacement  needed for the self-consistent solution of the Dyson equations  
\begin{align}
    \label{X(t)}
    \begin{split}
       \bar{X}_{\alpha } (t) =  & -\frac{2g}{\omega_{\textrm{ph}}} \big[ 2n_{\alpha}(0) - 1 \big] \\
+ 2g \int^{t}_0 d{\tau}& D^{R}_{0, \alpha \alpha} (t,{\tau}) \big[ n_{\alpha} ({\tau}) - n_{\alpha} (0) \big], 
    \end{split}
\end{align}
where the average electron occupation number is obtained from 
\begin{align} \label{n(t)}
n_{\alpha} (t)=\textrm{Im}\left[ G^{<}_{\alpha \alpha} (t,t) \right],
\end{align}
which is related to that in the original model as $n_{\alpha\uparrow}(t) + n_{\alpha\downarrow}(t) = 2n_{\alpha}(t)$. Notice that in these notations $n_\alpha=1$ means the double occupation of sublattice $\alpha$ (electron pairs). A substitution $2 n_\alpha \to n_\alpha$ recovers the form of Eq. (\ref{X(t)}) used in Ref. [\onlinecite{Schler2020}], where $n_\alpha=1$ corresponds to the half-filling.  

The occupation numbers as well as the phonon displacement are used to trace the dynamics of the metal-insulator transition. The insulating phase with the CDW is characterized by a generally non-symmetric occupation $n_0 \neq n_1$ as well as by the stationary lattice distortion manifested in the non-zero phonon displacement averages, $\bar{X}_0 = - \bar{X}_1 \neq 0$. The metallic phase does not have the lattice distortion and has no difference between the charge distribution, $\bar{X}_0 = \bar{X}_1 = 0$ and $n_0 =n_1 =0.5$. 

We also analyse the dynamics of the electron spectral function 
\begin{equation} \label{DOS}
    N(\omega,t)=-\frac{1}{\pi} \textrm{Im}\left[  \int_{0}^{t_\infty} d{\tau } e^{i\omega {\tau}} \sum_{\alpha} G^R_{\alpha \alpha}(t+ {\tau},t) \right],
\end{equation}
which give us the single-particle DOS and the gap (the choice of $t_\infty$ is dictated by the numerical efficiency). 

Equations (\ref{Dyson}) are solved numerically using solvers implemented in the  \textit{Non-Equilibrium Systems Simulation Library} (NESSI) \cite{Schler2020} which  were successfully used to study the dynamics of various correlated lattice systems, in particular, excited Mott insulators \cite{Eckstein2011, Ligges2018} and periodically driven cold atom systems  \cite{Sandholzer2019}. In this work we extended resources of this library to the case of the two site impurities and the matrix Dyson equations.

One notes that the initial Hamiltonian (\ref{H}) is degenerate with respect to the electron sublattice occupation. In order to obtain the temperature equilibrium CDW state with the broken symmetry, we adopt the method of quasi-averages \cite{BogolubovJr2014}. Following this method the symmetry is lowered initially by introducing a small difference between the sublattice chemical potentials. During a subsequent time evolution this symmetry breaking contribution is removed, however, the system evolves towards the equilibrium CDW state. The latter is used as the initial state to obtain the time evolution for the driven system.

\subsection{Material  parameters}

As noted above the hopping matrix elements of the TB model are extracted from the {\it ab initio} calculations. The effective matrix elements of the Bethe lattice are obtained using two conditions: 1) the width of the band is equal to that of the TB model $W\simeq 6$eV, 2) the shape of the DOS for the Bethe lattice model should be as close as possible to the TB model. This yields  $t_{\|}=1.0$eV and $t_{\perp}=0.5$eV for the intra- and inter-lattice hopping matrix elements.

Further, the phonon mode frequency is set to  $\omega_{\textrm{ph}}=70$me, which corresponds to the Raman breathing mode of $\text{BaBiO}_{\text{3}}$  \cite{Sugai1985, Tajima1992}. The strength of the electron-phonon coupling is estimated from the value of the gap in the single-particle DOS $\Delta \simeq 2.5$eV. This yields the coupling constant $g=0.35$eV which is close to $g=0.43$eV, reported earlier \cite{Khazraie2018}. In the calculations we consider  the case of zero doping with a half-filled band. The calculations are done at room temperature $T=290$K.

In what follows we investigate the dependence of the dynamics of the system on the frequency $\omega_0$ and intensity of the excitation laser pulse.  We consider a broad interval of optical frequencies $1{\rm eV} \lesssim \hbar \omega_0 \lesssim 4{\rm eV}$. For the excitation intensity we use the dimensionless parameter $A_0 = e a |E|/ (\omega_0 c \hbar)$, related to the field amplitude $|E|$, where  $a=4.34$\AA \, is the lattice constant of the BBO structure, $e$ is the elementary charge, $c$ is the speed of light. We consider relatively high intensities with the field amplitude $|E| \sim \Delta / e a $. This corresponds to the laser intensity $I_0 = |E|^2/Z_0 \simeq  10^{13} W/cm^{2}$, where $Z_0 = 376.73 \Omega$ is the vacuum impedance.

\begin{figure}
    \centering
    \includegraphics[height=8.0cm]{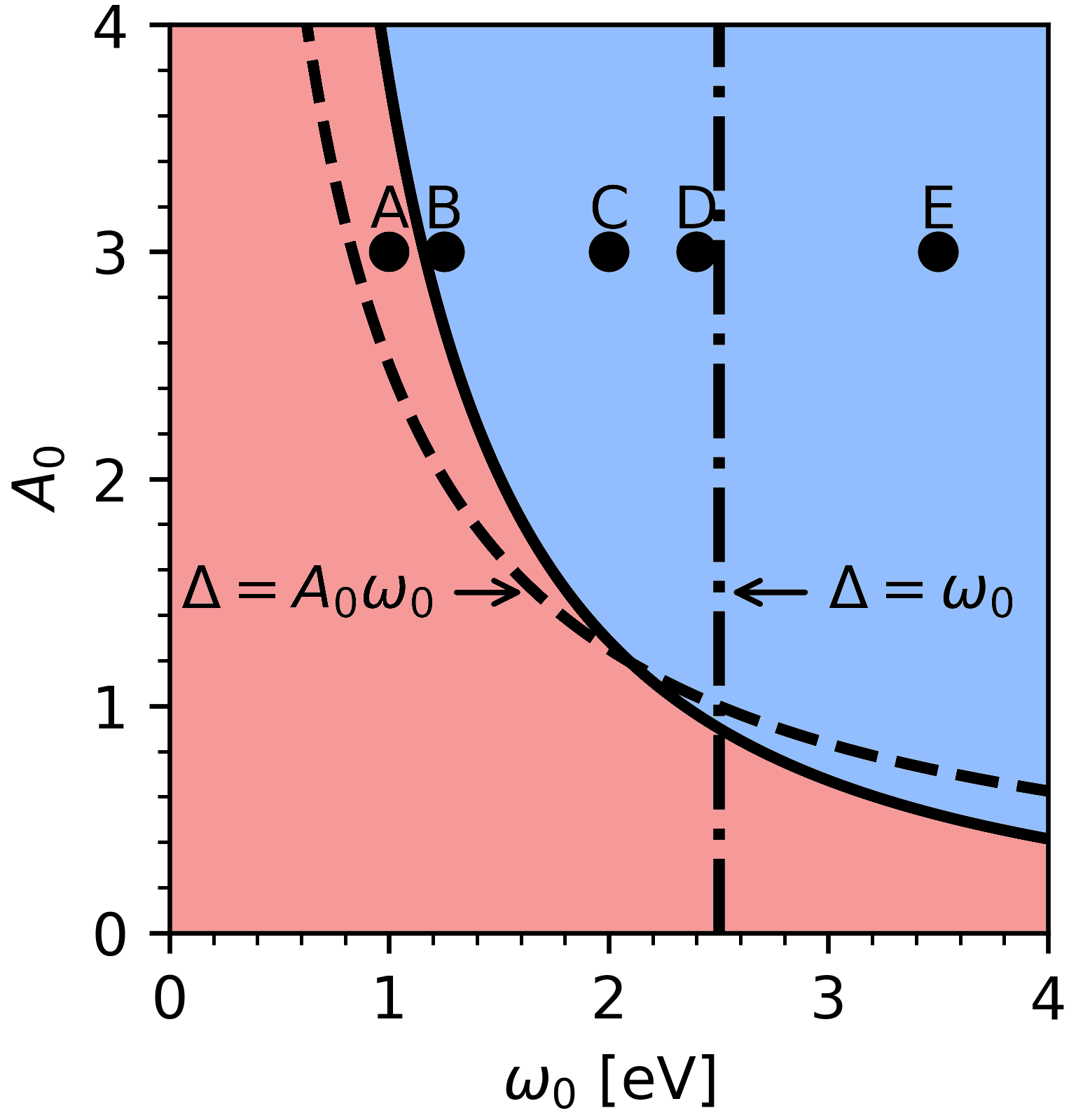}
    \caption{Phase diagram of types of the time evolution. Red colour marks the weak excitation domain, blue colour denotes the domain of strong excitation with the transient metal-insulator transition. The dashed line is given by $A_0 \omega_0 = \Delta$ ($= 2.5$eV) is an estimated threshold separating these regimes. The dashed-dotted line is given by $\hbar \omega_0 = \Delta$. Dynamics, calculated in points A-E is shown in Fig. \ref{fig:Table}. }
    \label{fig:PhaseDiagram}
\end{figure}

\begin{figure*}[t!]
\begin{center}
    \includegraphics[width=\textwidth]{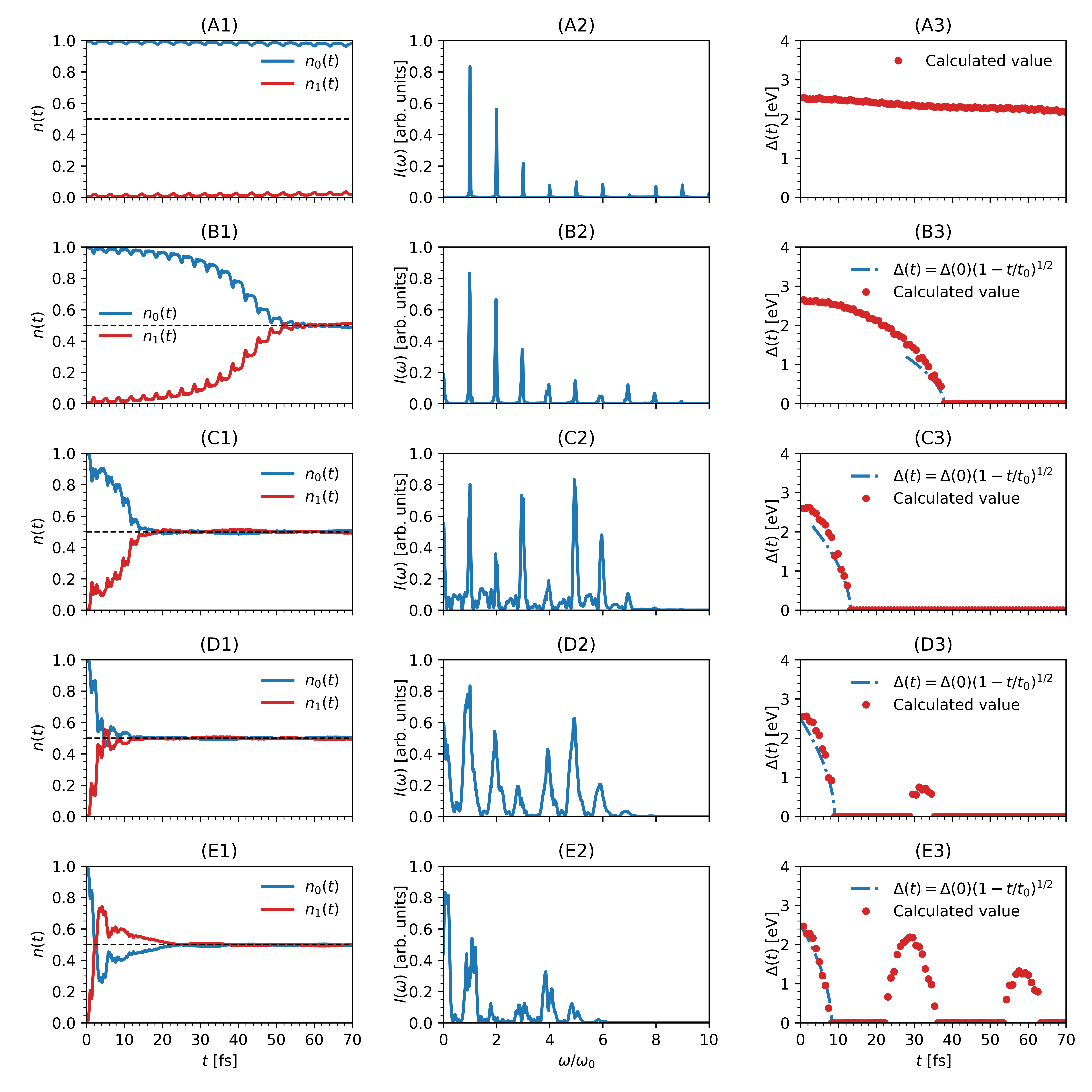}
\end{center}
\caption{Time evolution of the driven system calculated for parameters at points A-E in the diagram in Fig. \ref{fig:PhaseDiagram} is shown, correspondingly, in rows of Panels A-E. Panels A1-E1 plot the time evolution of the electronic occupation in sublattices $\alpha=0,1$. The second Panel column A2-E2 shows the spectra $I(\omega)$ of the electron dynamics as follows from the analysis of the Fourier component of the inter-lattice current flow. 
Panels A3-E3 plot the time evolution of the dielectric gap. The square root function $\Delta(t)=\Delta(0) (1-t/t_0)^{1/2}$ with  $\Delta(0)= 2.5$eV is used to fit the time dependence of the gap close to its vanishing point. }
\label{fig:Table}
\end{figure*}

\section{Dynamics of a driven system} 
\label{Results}

The dynamics of the system is investigated for the  continuous driving, switched on at $t=0$. The time evolution is obtained by solving the DMFT equations (\ref{Dyson})  briefly described above. Before the pulse arrival, at $t<0$, the system is in the equilibrium state, calculated self-consistently using the same formalism. 

The calculations are done by choosing the frequency $\omega_0$ and the amplitude $A_0$ of the driving pulse in the ranges shown in Fig. \ref{fig:PhaseDiagram}. From the complete time evolution we extract quantities that  characterize the transition from the insulating to the metallic state of the system: electronic occupation numbers $n_{\alpha}$ of the sublattice $\alpha$, which yield the electronic (pair) occupation at different octahedra $\text{BiO}_{\text{6}}$, the gap $\Delta(t)$ in the single-particle DOS, and the phonon displacement $X_\alpha(t)$ (octahedra distortion).

Before the arrival of the pulse the system is in the CDW state where the electronic pairs are shifted to one of the sublattice so that $n_0 =1$ and $n_1=0$, the gap in the DOS has a value $\Delta(0) = 2.5$eV and the octahedra are displaced with the amplitudes $X_0=0.42$\AA \, and $X_1=-0.42$\AA. 

\subsection{General remarks}

Before presenting numerical results it is instructive to discuss general factors that determine the  dynamical response of the system. 

When the excitation pulse is weak the dynamics is defined mostly by the linear response. The system has a gap in the single-particle DOS, equal to the bipolaron binding energy, and thus the response strongly depends on the relation between the excitation frequency and the gap. Here single-particle excitations take place when $\hbar \omega_0 > \Delta$.  Analogously to semiconductors the pulse creates electron-hole pairs, which in principle destroys the insulating state making the system conductive. However, the rate of such processes is small so that excited electron-hole pairs recombine back to the ground state and the system remains in the original CDW state with the gap. 

The situation becomes more complicated when the intensity of the laser pulse increases and multi-photon assisted tunnelling processes start to dominate the time evolution and the dynamics becomes  non-linear. When the driving is very strong one can reach a situation where the electronic structure of the system changes so strongly that the lattice distortion disappears together with the gap resulting in the transient metal-insulator transition. One can roughly estimate the onset of such processes as when the field reaches such strength that it overweights the gap-related potential barrier between the different octahedra. Then electrons start moving freely between the sublattices thereby eliminating their asymmetric distribution, lattice static distortion and the spectral gap. This threshold field amplitude is estimated as $e E a \simeq \Delta$, which yields the field strength interval chosen in this work. 

The threshold of the driving amplitude given by $A_0  \omega_0 \simeq \Delta$ as well as the resonance line $\Delta = \hbar \omega_0$  are shown by the dash-dotted and dashed line in the phase diagram of Fig. \ref{fig:PhaseDiagram} defining the domains of the dynamics types. In this work we are interested in the regime of stronger driving where the CDW state of the system can be altered. 

 \subsection{Dynamics of the electronic subsystem}

Results of the numerical calculations reveal several types of dynamics patterns. Depending of the final state of the system these are divided into two large classes, which occur in the coloured domains in Fig. \ref{fig:PhaseDiagram}. In the regime we call "weak driving"  (red domain) the evolution is slow an the system keeps its original insulating state even at very large times. In the regime of "strong driving" (blue domain) the system state rapidly changes to the gapless metallic state. These types interchange rather abruptly when passing the vicinity of the solid black crossover line in Fig. \ref{fig:PhaseDiagram}. This line represents a summary of the numerical results, however, it is surprisingly close to the simple estimate above, shown by the dashed line in Fig. \ref{fig:PhaseDiagram}.

Figure \ref{fig:Table} shows the time evolution calculated for several representative points A-E, shown in the diagram in Fig. \ref{fig:PhaseDiagram}. These points are selected to illustrate typical dynamical patterns in both the weak and the strong driving regimes. We  have done the calculations also for many other points in this diagram and they demonstrated very similar results. An interested reader can find the time evolution of the system for additional points on the diagram in the Supplemental Material \cite{Suppl}.

Panels (A1-E1) in Fig. \ref{fig:Table} plot the time evolution of the occupation $n_\alpha$ for sublattices $\alpha = 0,1$. Panels (A2-E2) show the spectrum of the dynamics, calculated as the frequency dependent absolute value of the Fourier component of the inter-lattice current flow $I(\omega) \sim \left| \textbf{\emph{j}}(\omega) \right|^2$, where $j(\omega)$ is the Fourier transform of the current $j(t) = \dot n_1(t) - \dot n_0(t)$. Finally, Panels (A3-E3) plot the time evolution of the gap in the single particle spectrum.  

\begin{figure}[t!]
\begin{center}
    \includegraphics[width=0.5\textwidth]{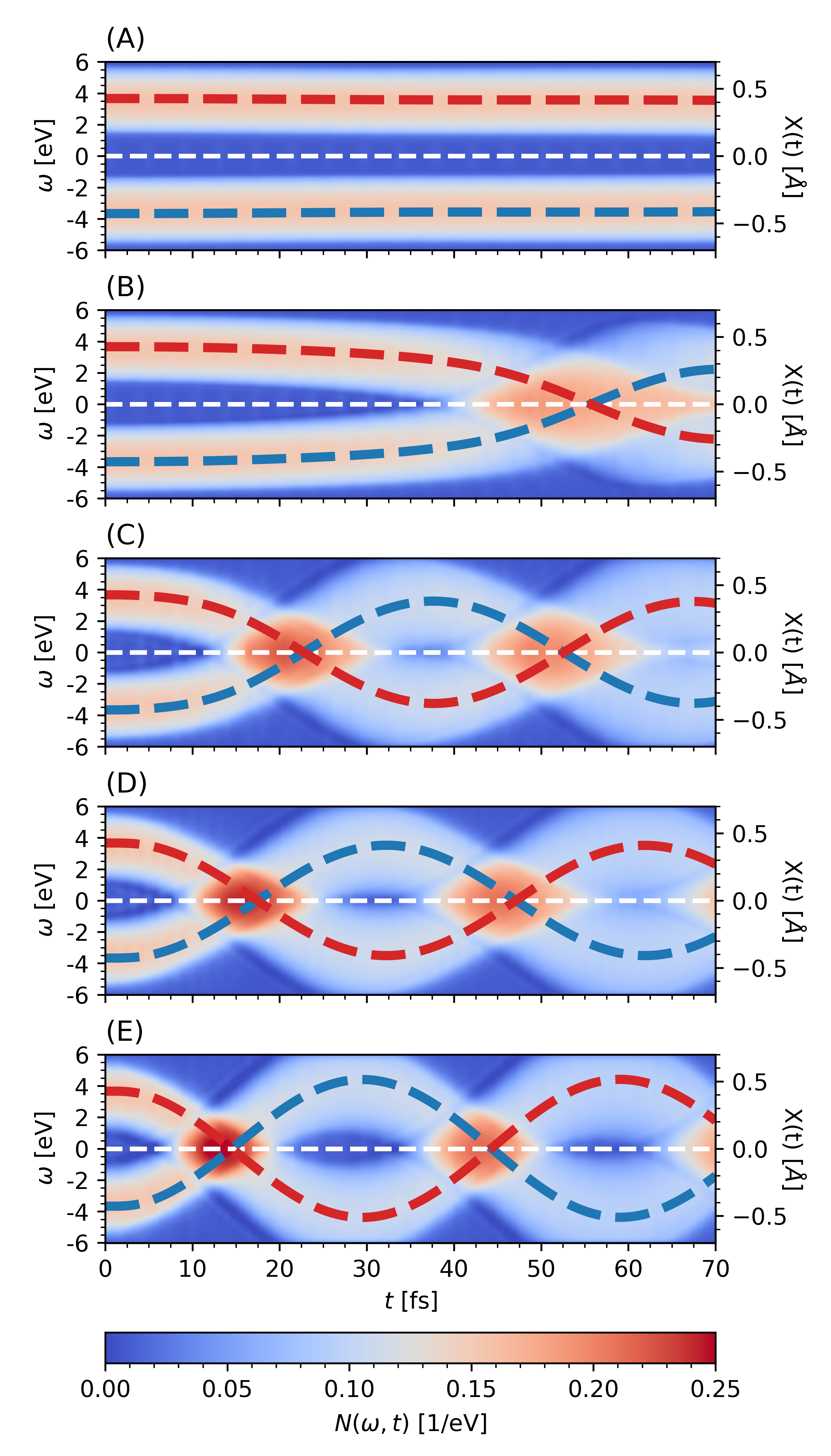}
\end{center}
\caption{Time evolution of the electron DOS, shown as the colour density plot, and the phonon-induced deformation of the sublattices, shown by two dashed lines. The blue dashed line shows $X_0(t)$, and the red $X_1(t)$. Panels (A-E) correspond to points A-E in Fig. \ref{fig:PhaseDiagram}. }
\label{fig:DOS_X}
\end{figure}

At point A [Fig. \ref{fig:PhaseDiagram}] the system is in the weak driving regime manifested by the weak time dependence of the electron occupation in Panel (A1). The corresponding changes in gap, shown in Panel (A3), are also small, $\Delta(t)$ saturates when the time increases and shows no sign of abating. At the same time the spectra of the time evolution in Panel (A2) reveals that the dynamics comprises a large non-linear component with the noticeable contribution of the high harmonics even to the order $n=10$. Appearance of high harmonics in this system is similar to that reported earlier for a strongly driven Mott insulator \cite{Murakami2018}. We note in passing that the name "weak driving" for this regime means only that the insulating state does not change in the evolution. 

Sharp changes of the dynamical pattern when the system crosses into the strongly driving regime (blue colour domain in Fig. \ref{fig:PhaseDiagram}) are illustrated by the results obtained at point B. Although B is close to A the electronic occupation now rapidly decays on the timescale of 50fs [Panel (B1)].  Furthermore, the occupation rapidly approaches a new quasi-equilibrium state that has equal electronic occupations, $n_0=n_1=0.5$, indicating that the CDW is destroyed. This is also seen in Panel (B3) which demonstrates that the gap disappears completely after $t_0 \simeq 40$fs. It is worth noting that the gap disappears with time abruptly. Its time evolution follows not a gradual decay but approaches zero as a square root dependence $(1-t/t_0)^{1/2}$, reminiscent of a time dependent phase transition. 

The spectrum of the occupation dynamics in Panel (B2) demonstrates a large contribution of high harmonics. It is worth noting, however, that the relative weight of the higher harmonics decreases in comparison with point A [Panel (A2)]. Deeper in the strong driving regime (point C in Fig. \ref{fig:PhaseDiagram}) the dynamics remains qualitatively similar to that at point B. Nevertheless, the electronic occupation relaxes faster to its new quasi-equilibrium state, the characteristic time for the gap disappearance decreases to $t_0 \simeq 15$fs. One notices a much longer oscillations in the time evolution of the occupation in Panel (C1). These are related to phonons, which is discussed below.  It is important to note that point B is still substantially below the resonance line $\hbar \omega_0 \ll \Delta$ and, therefore, the dynamics, at least initially till the gap disappearing, is mainly defined by the tunnelling through the gap related potential barrier.   

The point D is deeper in the strong driving domain, approaching the resonance condition $\hbar \omega_0 = \Delta$. Strictly speaking, this resonance corresponds to the single-photon picture in the linear response, whereas in our case the dynamics is highly non-linear and involve many-photon transitions. Nevertheless, our calculations demonstrate noticeable changes in the dynamics in the vicinity of this line, most likely because of the much increased contributions of the single-particle excitation processes in the initial stage of the time evolution. 

The proximity to the resonance $\hbar \omega_0 \simeq \Delta$ is manifested in a sharp decrease of the decay time in the electron occupation dynamics [Panel (D1)],  accompanied by a similarly fast disappearance of the gap, $t_0 \simeq 10$fs [Panel (D3)]. Remarkably, the gap reappears after $t_1 \simeq 30$fs.  We stress that such reappearance does not take place at point C even though it is relatively close to D, the longer time calculations demonstrate absence of the gap in this point.

Qualitative changes in the dynamics at point D, in contrast to C, are also manifested in that its spectrum, shown in Panel (D2), becomes essentially continuous (although the peaks at the integer harmonics are still visible). The spectrum reveals that the dynamics now has a significant contribution of the slow time evolution with $\omega \sim 0$, which is related to the dynamics of phonons and of the gap reappearance. 

With a further increase in the frequency direct excitations over the gap become more and more significant. At point E, the frequency is so high that the laser field moves both electrons at the same time leading to the "overshoot" in the densities, $n_0(t)< n_1(t)$ [Panel (E1)]. The effect of the gap reappearance becomes more pronounced [Panel (E3)], the gap reappears several times before vanishing completely. In contrast the long time calculations at point D reveal only a single gap reappearance. The spectrum for point E, shown in Panel (E2), becomes "more continuous": the integer harmonics peaks are less pronounced.  The higher frequency contribution to the spectrum is suppressed comparing to points A-D. 

 We note that the same results are obtained when the interchange between the weak and strong driving domain goes along other lines in the phase diagram in Fig. \ref{fig:PhaseDiagram}. The  the crossover line actually is obtained as a summary of the crossing points at which the dynamic changes. As an example the Supplemental material \cite{Suppl} demonstrates changes in the time evolution along the lines $A_0 = 2$ and $A_0 = 1$, as well as along the vertical line $\omega_0 = 3.5$eV.

\begin{figure}[t!]
\begin{center}
    \includegraphics[width=0.5\textwidth]{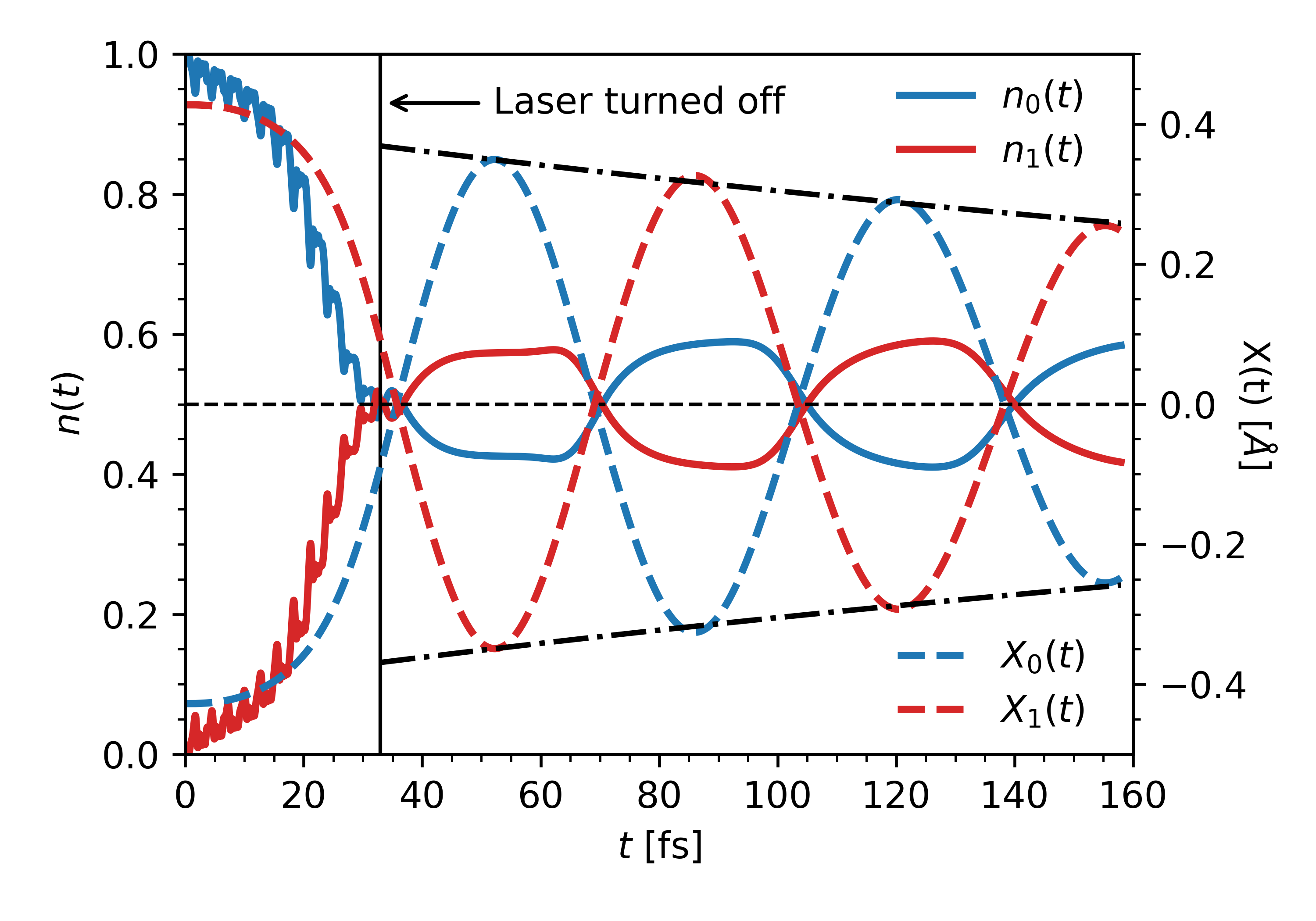}
\end{center}
\caption{The dynamical response of the system at point C in Fig. \ref{fig:PhaseDiagram}), the laser field is turned off at the phase transition $t_0=33$fs. Time evolution of the occupation of sites '0' and '1'  is shown by solid lines, lattice displacements are represented by the dashed lines, the  dash-dotted line gives the decaying envelope of the  phonon displacement. }
\label{fig:LaserOff}
\end{figure}

 \subsection{Spectral gap and deformation dynamics}

A further insight into the system dynamics is obtained by comparing the time evolution of the electronic degrees of freedom with that of the lattice deformation. The colour density plot in Fig.  \ref{fig:DOS_X} shows the time dependence of the electronic DOS superimposed with the deformation amplitude $X_\alpha(t)$ mediated by the phonon oscillations. As expected for the CDW state, the dielectric gap in the DOS closely follows the lattice distortion, disappearing precisely at points where it vanishes, $X_\alpha =0$. 

The gap reappearance decays on the time scale of $\sim 100$fs, whereas the phonon oscillations do not decay because the driving constantly pumps energy into the system. The phonon decay sets in when the driving is switched off. It has to be noted that the Holstein model does not account for mechanisms for the phonon relaxation, and thus phonons decay only via their coupling to electrons. It should also be noted that in this work we are in the limit of very strong phonon-electron coupling. In this case we are in the limit of the reduced phonon relaxations in comparison to that for electrons, as were reported earlier for the Holstein model \cite{Murakami2015}. 

This is shown in Fig. \ref{fig:LaserOff}, where the laser destroys the CDW state and then is turned off at $t=33$fs. The phonon dynamics then induces oscillations in the electronic occupations, which decay gradually with the characteristic time of $\tau_{ph} = 350$fs. However, the new equilibrium state is not the original insulating CDW but is the metallic state with $n_0 = n_1=0.5$ without the lattice distortion.  

\section{Conclusions} 
\label{Conclusions}

This work studied ultra-fast dynamics of undoped $\text{BaBiO}_{\text{3}}$, excited by a strong continuous laser pulse, with a goal to investigate the laser-induced transition between insulating and metallic phases.  The material is described by the Holstein model. The dynamics is obtained using the approach that combines the DMFT and the self-consistent Migdal approximation for the self-energy. The method takes into account quantum coherences and allows to calculate time evolution on the ultra-fast time scales.  To simplify the calculations the tight-binding model for the electronic states is mapped on the double Bethe lattice model, for which the DMFT equations are greatly simplified. 

The dynamics was calculated in a wide range of optical frequencies and excitation intensities. The calculations demonstrated two qualitatively different regimes: the weak driving regime, where the insulating phase remains intact, and the strong driving one, in which the system undergoes a transient change into the metallic state without CDW and lattice deformations.  The transition between the two regimes is reliably estimated from the condition the driving field overcomes the gap in the single particle DOS. 

The insulating state disappears abruptly: the gap vanishes with time by following the square root law, typical for phase transitions. The transition is of purely quantum nature, it is caused not by the temperature, but by the coherent driven dynamics of electron and phonon degrees of freedom. It is not connected to the nonlinearity: in both cases the dynamics can be highly nonlinear producing a large number of higher harmonics.   

Typical characteristic times of the relaxations in the electronic subsystem and that of the insulator-metal transition are found to be in the range of tens of fs, considerably smaller than the characteristic phonon times. Thus the metallic state is achieved on the background of strongly oscillating lattice. However, those oscillations were found to affect the electronic states in a surprising way, causing a periodical reappearance of the gap. This takes place when the driving frequency exceeds the gap. The reappearance time is locked to the period of the lattice oscillations, although the amplitude of the reappearance decays much faster. We also noted that when the laser frequency exceeds the gap the spectrum of the dynamics becomes continuous, although the peaks related to the higher harmonics remain. In this case the spectrum of the electronic dynamics acquires a significant low frequency component, related to the gap reappearance. 

Although the adopted model is simplified it is frequently used in the studies of similar structures and is expected to capture the essential physics of the metal-insulator transition, especially on the ultra-fast time scales where other relaxation processes are not important. The model parameters were chosen to match those of the real material. In the calculations we have exploited experimentally available laser frequencies and intensities. We also found that the appearance of different regimes can be understood using simple intuitive estimations, which are, nevertheless, quantitatively not very far  from those obtained by the calculations, speaking in favor that the results can describe real dynamics and the transient metal-insulator transition in  strongly driven $\text{BaBiO}_{\text{3}}$ compounds.

\section*{Acknowledgements}
The work was supported by the Russian Foundation for Basic Research under the Project 18-02-40001 mega. 
Y.Z. also thanks for support the Deutsche Forschungsgemeinschaft under the Project-ID 314695032 – SFB 1277.

\bibliography{bibliography}

\end{document}



\title{Laser-induced ultrafast insulator--metal transition in  $\text{BaBiO}_{3}$\\ Supplemental material}

\author{Alexander E. Lukyanov$^{1,2}$}
\author{Vyacheslav D. Neverov$^{1,2}$}
\author{Yaroslav V. Zhumagulov$^{3,2,1}$}
\author{Alexey P. Menushenkov$^{1}$}
\author{Andrey V. Krasavin$^{1}$}
\author{Alexei Vagov$^{4,2}$}

\affiliation{$^{1}$National Research Nuclear University MEPhI, Kashirskoye shosse 31, Moscow, 115409, Russian Federation}
\affiliation{$^{2}$ITMO University, St. Petersburg 197101, Russia}
\affiliation{$^{3}$University of Regensburg, Regensburg, 93040, Germany}
\affiliation{$^{4}$Institute for Theoretical Physics III, University of Bayreuth, Bayreuth 95440, Germany}

\date{\today}

\maketitle

\begin{figure}
    \centering
    \includegraphics[width=0.8\columnwidth]{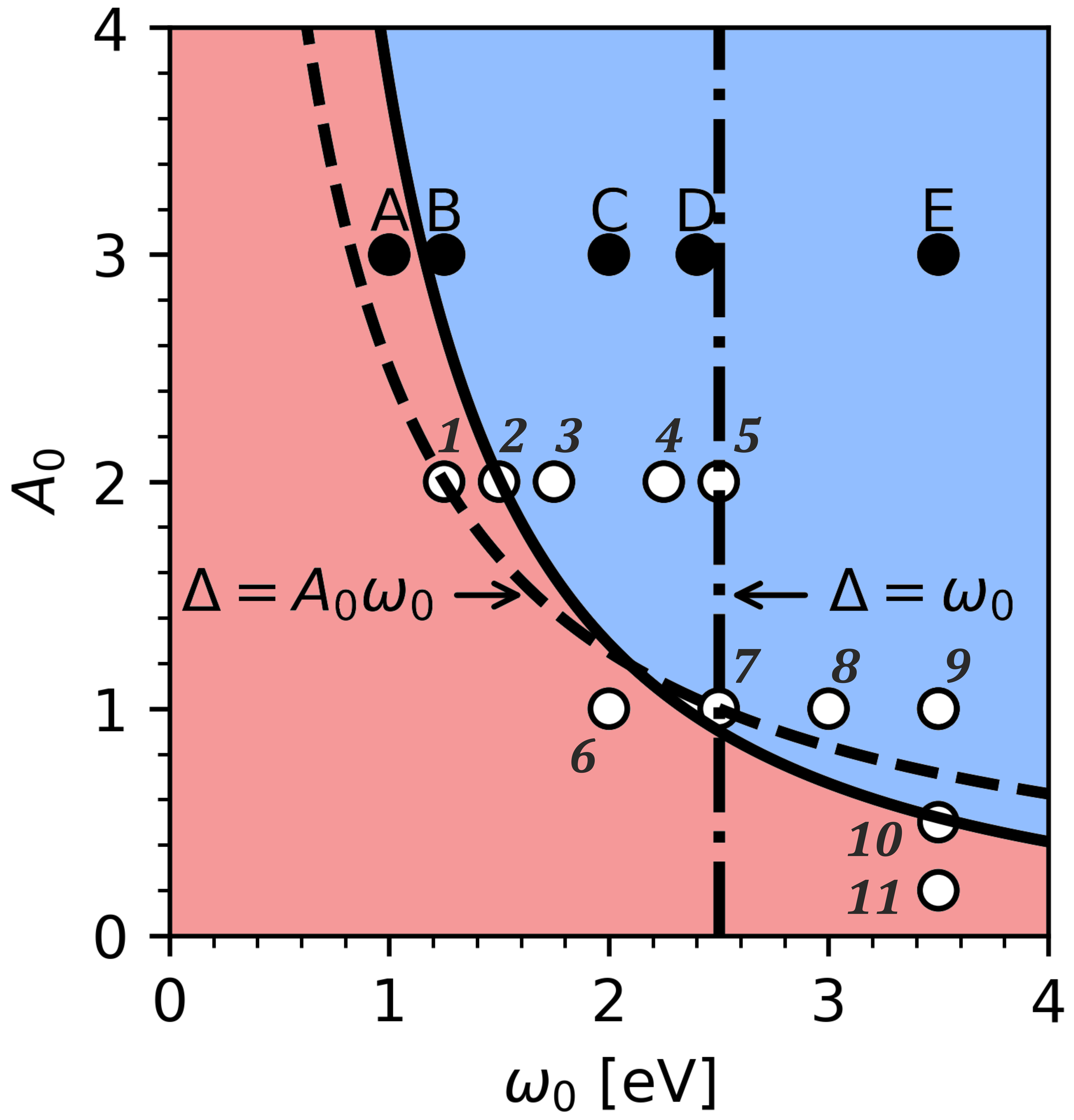}
    \caption{Phase diagram of types of the dynamics, as in Fig. 5 of the MT. Red colour marks the regime of the weak driving regimes, blue colour denotes the domain of the strong driving with the transient metal--insulator transition (see the MT), the black continuous line is the crossover between the two regimes. The dynamics calculated at black points $A-E$ is shown in Fig. 6 of the MT. The dynamics at additional white points $1-11$ is shown in Figs. \ref{fig:S2}, \ref{fig:S3} and \ref{fig:S4} of the Supplemental material.}
    \label{fig:PD_Supp}
\end{figure}

Here the dynamics of the driven system calculated at additional points $1-11$ of the phase diagram in Fig. \ref{fig:PD_Supp}. This amend the calculations shown in Fig. 6 of the main text (MT), obtained at black points $A-E$ in Fig. \ref{fig:PD_Supp}. Results for the additional points support the main finding of the work that the phase space in the $\omega_0 - A_0$ plane is divided into two domains of qualitatively different dynamical patterns, marked by red and blue colour in Fig. \ref{fig:PD_Supp}. In the red domain the system is in the regime of the weak driving, where the insulating phase with the gap in the single-particle spectrum survives. The blue domain is the strong driving regime, where the system reaches the metallic gapless state on the ultra-fast time scale. The dynamics type changes rapidly when the system crosses the border between the domains, illustrated by the black continuous line in Fig. \ref{fig:PD_Supp}.

\begin{figure}
\includegraphics[width=7.5cm]{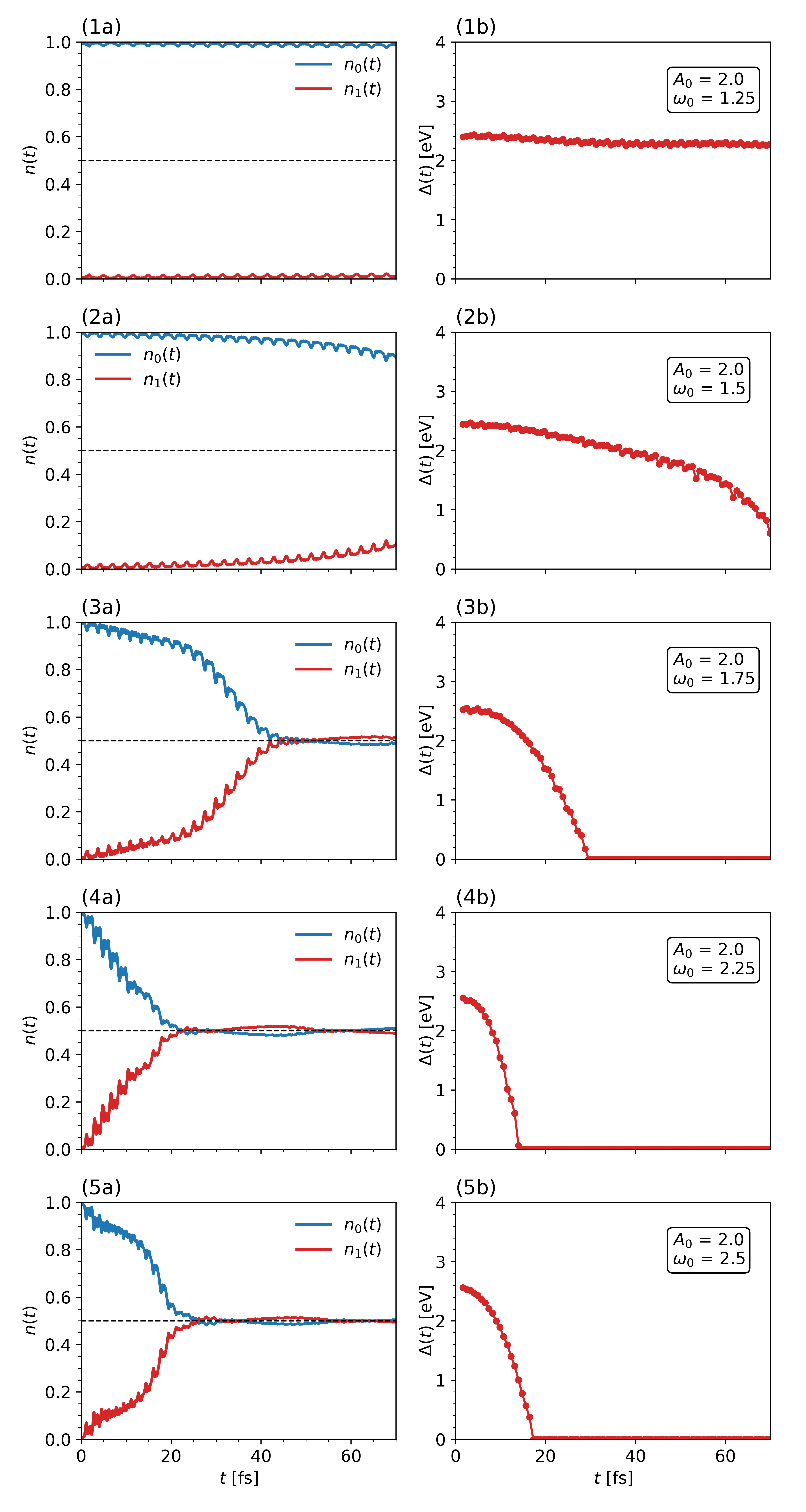}
\caption{Lattice occupations $n_{0;1}(t)$ (left panels) and dielectric gap $\Delta(t)$ (right panels) calculated at points $1-5$ in Fig. \ref{fig:PD_Supp}.} 
\label{fig:S2}
\end{figure}

Figure \ref{fig:S2} shows how the time dynamics changes when the system crosses the line between the domains at $A_0 = 2$ (lower than for points $A-E$). Point 1 ($\omega_0=1.25$eV) is at the weak driving regime, where the insulating state is not broken. It is clearly seen  in Figs. \ref{fig:S2} (1a) - (1b), which show sublattice occupations $n_i(t)$ and gap $\Delta(t)$, respectively. Figures \ref{fig:S2} (2a) - (2b) demonstrate the dynamics at point 2 ($\omega_0=1.5$eV), located close to the crossover to the strong driving regime. At this point the system eventually approaches the metallic equilibrium state ($n_0 = n_1 =0.5$ and $\Delta =0$) on the time scale of $t \simeq 100$fs. At point 3 ($\omega_0=1.75$eV), deeper in the strong driving regime, the characteristic time scale for breaking the insulating phase shortens to $t\simeq 40$fs [Figs. \ref{fig:S2} (3a) - (3b)]. These crossover in the dynamical pattern are the same as observed at points $A-C$ and shown in Figs. 6 (A) - (C) in the MT.

Deeper inside the strong driving regime, i.e. when the driving frequency $\omega_0$ increases, the characteristic time scale continues to decrease, reaching the minimum in the vicinity of the resonance line $\omega_0 = \Delta$  at point 4 ($\omega_0 = 2.25$eV) [Figs. \ref{fig:S2} (4a) - (4b)]. At larger $\omega_0$ the time scale increases slightly [Figs. \ref{fig:S2} (5a) - (5b)]. The results for obtained for points $1-5$ in Fig.  \ref{fig:S2} are thus similar to those at points $A-E$, shown in Fig. 6 in the MT. 

\begin{figure}[H]
\includegraphics[width=7.5cm]{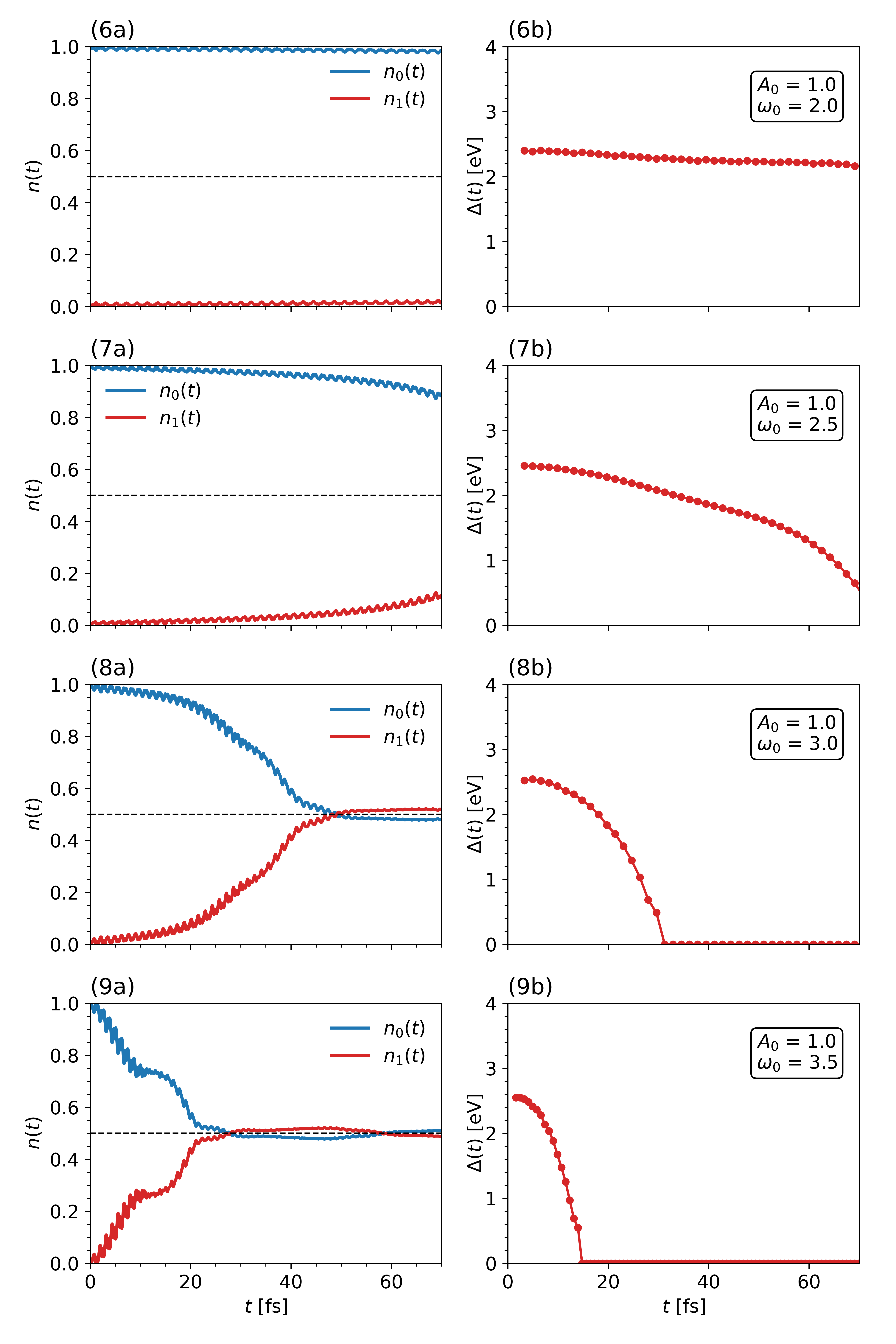}
\caption{Lattice occupations $n_{0;1}(t)$ (left panels) and dielectric gap $\Delta(t)$ (right panels) calculated at points $6-9$ in Fig. \ref{fig:PD_Supp}.}
\label{fig:S3}
\end{figure}

Similar results are also obtained at lower driving $A_0 = 1$, at points $6-9$ [Fig. \ref{fig:S3}].  At point 6 ($\omega_0 = 2.0$eV) one observes the weak driving dynamical picture [Fig. \ref{fig:S3} (6a) - (6b)], that changes to the strong driving regime close to  point 7 [Fig. \ref{fig:S3} (7a) - (7b)]. This point is also located at the resonance (dashed dotted line) which is not as pronounced as when the driving is stronger (at points 4 and 5, and points C and D). Instead one sees a decrease in the characteristic time to $t\simeq 40$fs at point 8 ($\omega_0 = 3$eV) [Fig. \ref{fig:S3} (8a) - (8b)] and then to $t\simeq 20$fs at point 9  ($\omega_0 = 3.5$eV)   [Fig. \ref{fig:S3} (9a) - (9b)]. 

\begin{figure}[H]
\includegraphics[width=7.5cm]{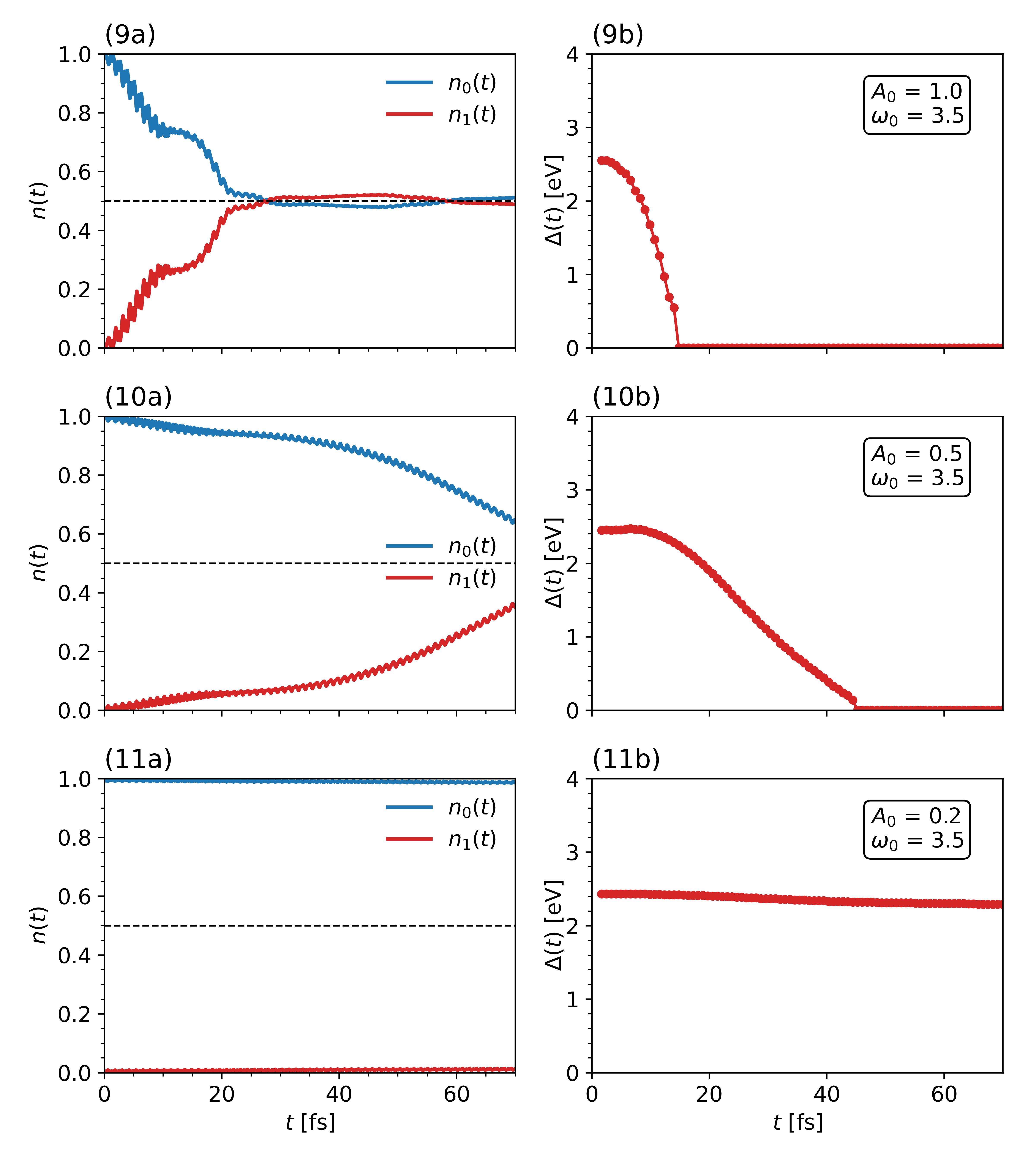}
\caption{Lattice occupations $n_{0;1}(t)$ (left panels) and dielectric gap $\Delta(t)$ (right panels) calculated at points $9-11$ in Fig. \ref{fig:PD_Supp}.}
\label{fig:S4}
\end{figure}

Finally, we demonstrate the crossover between the regimes of the weak and the strong driving by crossing the crossover line vertically. This is illustrated in Fig. \ref{fig:S4} that shows results for points $9-11$ with $\omega_0 = 3.5$eV. Here one sees similar changes in the dynamics, starting from point $11$ ($A_0 = 0.2$) in the domain of weak driving [Fig. \ref{fig:S4} (11a) - (11b)], going to the strong driving regime at point $10$ ($A_0 = 0.5$)  [Fig. \ref{fig:S4} (10a) - (10b)] and then to point $9$ ($A_0 = 1$) [Fig. \ref{fig:S4} (9a) - (9b)].

